\documentclass[twocolumn,twoside]{IEEEtran}

\usepackage{mathpazo}
\usepackage{times}
\usepackage{amsmath}
\usepackage{amsfonts}
\usepackage{latexsym}
\usepackage{amssymb}
\usepackage{cite}

\usepackage{upref}
\usepackage{theorem}
\usepackage{graphicx}
\usepackage{psfrag}
\usepackage{color}

\usepackage[finalnew]{trackchanges}
\usepackage{setspace}

\newcommand\isitremove[1]{}

\addeditor{alex}
\newcommand\alex[1]{\add[alex]{#1}}
\newcommand\alexr[1]{\remove[alex]{#1}}

\newtheorem{alg}{Algorithm}
\newtheorem{claim}{Claim}
\newtheorem{theorem}{Theorem}
\newtheorem{lem}{Lemma}
\newtheorem{clm}[lem]{Claim}
\newtheorem{definition}{Definition}

\newtheorem{coro}{Corollary}

\newcommand{\BA}{\begin{alg}} \newcommand{\EA}{\end{alg}}
\newcommand{\BE}{\begin{enumerate}} \newcommand{\EE}{\end{enumerate}}
\newcommand{\BT}{\begin{theorem}} \newcommand{\ET}{\end{theorem}}
\newcommand{\BL}{\begin{lem}} \newcommand{\EL}{\end{lem}}
\newcommand{\BCM}{\begin{clm}} \newcommand{\ECM}{\end{clm}}
\newcommand{\BCR}{\begin{coro}} \newcommand{\ECR}{\end{coro}}

\newcommand{\BI}{\begin{itemize}} \newcommand{\EI}{\end{itemize}}

\def\FullBox{\hbox{\vrule width 8pt height 8pt depth 0pt}}
\newcommand{\qed}{\;\;\;\FullBox}
\newenvironment{prf}{\noindent{\bf Proof:~~}}{\(\qed\)}
\newcommand{\BPF}{\begin{prf}} \newcommand {\EPF}{\end{prf}}
\newenvironment{proofof}[1]{\noindent{\bf Proof of {#1}.~}}{\endprf}
\newcommand{\BPFOF}{\begin{proofof}} \newcommand {\EPFOF}{\end{proofof}}

\newcommand{\eat}[1]{}

\begin{document}

\title{Group Testing on General Set-Systems}
\author{\large
Mira Gonen\thanks{Mira Gonen is with the Department of Computer Science, Ariel University, Ariel, 40700, Israel (e-mail: mirag@ariel.ac.il).},
Michael Langberg\thanks{Michael Langberg is with the Department
  of Electrical Engineering, State University of New-York at Buffalo, Buffalo, NY 14260, USA (e-mail: mikel@buffalo.edu). Work supported in part by NSF grant 1909451.}, and
Alex Sprintson\thanks{Alex Sprintson is with the Department of Electrical and Computer Engineering, Texas A\&M University, College Station, TX 77843-3128, USA (e-mail:  spalex@tamu.edu). }
  %
}
  %


\maketitle

\thispagestyle{empty}

\begin{abstract}

Group testing is one of the fundamental problems in coding theory and combinatorics in which one is to identify a subset of contaminated items from a given ground set.
There has been renewed interest in group testing recently due to its applications in 
\alex{diagnostic virology, including}
pool testing for the novel coronavirus.
The majority of existing works on group testing focus on the
\emph{uniform} setting in which any subset of size $d$ from a ground set $V$ of size $n$ is potentially contaminated.

In this work, we consider a {\em generalized} version of group testing with an arbitrary set-system of potentially contaminated sets.
The generalized problem is characterized by a hypergraph $H=(V,E)$,
where $V$ represents the ground set and edges $e\in E$ represent potentially contaminated sets.
The problem of generalized group testing is motivated by practical settings in which not all subsets of a given size $d$ may be potentially contaminated,  rather, due to social dynamics, geographical limitations, or other considerations, there exist subsets that can be readily ruled out.
For example, in the context of pool testing, the edge set $E$ may consist of families, work teams, or students in a classroom, i.e., subsets likely to be mutually contaminated.
The goal in studying the generalized setting is to leverage the additional knowledge characterized by $H=(V,E)$ to significantly reduce the number of required tests. 

The paper considers both adaptive and non-adaptive group testing and makes the following contributions. First, for the non-adaptive setting, we show that finding an optimal  solution for the generalized version of group testing is NP-hard. For this setting, we present a solution that requires $O(d\log{|E|})$ tests, where $d$ is the maximum size of a set $e \in E$. Our solutions generalize those given for the traditional setting and are shown to be of order-optimal size $O(\log{|E|})$ for hypergraphs with edges that have ``large'' symmetric differences.
For the adaptive setting, when edges in $E$ are  
of size exactly $d$, we present a solution of size $O(\log{|E|}+d\log^2{d})$ that comes close to the lower bound of $\Omega(\log{|E|} + d)$.





\end{abstract}

\section{Introduction}






Group testing is one of the fundamental problems in coding theory, statistical inference, and combinatorics due to its practical importance in a broad range of  applications, such as multi-access communication \cite{1057026}, pattern matching \cite{10.1016/j.jcss.2009.06.002}, molecular biology, and others. The problem has deep connection to other fundamental problems in combinatorics and coding theory \cite{5967914}.

The group testing problem was subject to a large number of studies since its introduction more than 75 years ago (refer to, e.g.,\cite{DBLP:journals/corr/abs-1902-06002,dh00} and references therein). 
An instance of the group testing problem includes a ground set $V$ of items of size $n$,
a subset of which may be \emph{contaminated}
(we refer to the latter set
as a \emph{contaminated set}).
In the traditional setting, for a given parameter $d$, any subset of $V$ of size $d$ may potentially be a contaminated set.
The contaminated set can be detected through a pooling process which includes a series of tests, where each test reveals the existence of a contaminated item in the tested subset of items. The goal of the group testing problem is to design a minimum set of tests that can identify the contaminated set of items in $V$. The testing algorithms can be constructed in a non-adaptive manner, i.e., fixed in advance, or in an \emph{adaptive} manner, i.e., each test can depend on the outcome of previous tests.

In this work, we study a \alex{more general} version of group testing, termed here {\em generalized group testing}, in which the set-system of potentially contaminated sets is characterized by a hypergraph $H=(V,E)$, in which the vertex set 
$V$ represents the ground set and the hyperedges in $E$ represent potentially contaminated sets.
Traditional group testing thus corresponds to the hypergraph $H$ with edge set $E$ consisting of {\em all} subsets of $V$ of size $d$.
Our study of arbitrary edge sets $E$ grants the flexibility required for a broad range of settings, including those in which the set-system $E$ of potentially contaminated sets does not have any uniformity or regularity properties. For example, in the context of pool testing, $E$ can represent potentially contaminated sets that correspond to families, work teams, or groups of friends or students in a classroom \cite{pmlr-v130-nikolopoulos21a}.
The goal in studying the generalized setting is to leverage the additional knowledge characterized by $H=(V,E)$ to minimize the number of required tests.


\textbf{Contribution.} The paper considers both adaptive and non-adaptive group testing and makes the following contributions. First, for the non-adaptive setting, we show that finding an optimal 
solution for the generalized version of the group testing problem is NP-hard, and approximating the optimal solution within a factor of $1+\varepsilon$ (for sufficiently small $\varepsilon$) is as hard as coloring a 3-colorable graph with $n^{\varepsilon}$ colors.
The latter is a well known open problem, e.g.,~\cite{KLS00,langberg2008graph}.
For the non-adaptive setting, we present a solution that requires $O(d\log{|E|})$ tests, where $d$ is the maximum size of a set $e \in E$. Our solutions generalize those given in the classical setting and are shown to be of order-optimal size $O(\log{|E|})$ for hypergraphs with edges that have ``large'' symmetric differences.
For the adaptive setting, in which all edges in $E$ are  of size exactly $d$, we present a solution of size $O(\log{|E|}+d\log^2{d})$ that comes close to the lower bound of $\Omega(\log{|E|} + d)$.
For the adaptive setting in which $d$ is the maximum size of a set $e \in E$, we obtain an upper-bound of
$O(\log{|E|}+d^2)$.



\textbf{Related works.} The overwhelming  majority of works on  group testing fall under the traditional setting in which $H=(V,E)$ includes all edges of size $d$ (or, alternatively, of size at most $d$).
Known upper and lower bounds in this context are reviewed in Section~\ref{sec:bounds}.
For hypergraphs $H$ that differ from the traditional setting, less is known.
Using our notation, Nikolopoulos et al. \cite{pmlr-v130-nikolopoulos21a} study hypergraphs  $H=(V,E)$ that have a certain \emph{community structure}. Specifically, \cite{pmlr-v130-nikolopoulos21a}  assumes that $V$ consists of $F$ disjoint groups (referred to as \emph{families}) and studies the special setting in which $E$ includes all edges that intersect a bounded number of families.
The paper leverages the structure of $H=(V,E)$ to maximize the efficiency of the group testing algorithms in the adaptive, non-adaptive, and probabilistic settings. A few related papers \cite{zhu2020noisy,9414034} focus on leveraging side-information (e.g., that can be obtained from contact tracing) to make the decoding algorithm faster.
Finally, {\em graph-constrained group testing}, a variant of the group testing problem
where the {\em tests} must conform to constraints imposed by a graph, is considered, e.g., in~\cite{ckms10,kz12,lmms19}.
The results of this work differ significantly from those above as we study general set-systems of potentially contaminated sets, and do not place any  constraints on the tests used.


\section{Problem Formulation}
\label{sec:prelim}


An instance of the group testing problem includes a ground set $V$  of $n$ items,
a subset of which may be contaminated, with the goal of
designing a minimum set of tests that can identify the contaminated items.
We first define traditional non-adaptive group testing.

\begin{definition}[traditional group testing (non-adaptive)]
\label{def:trad_gt}
For a ground set $V$ of size $n$ and a parameter $d$, find a minimum size family ${\cal T}$ of subsets of $V$ such that for any $A,B\subseteq V$ of size $d$ there exists $T\in {\cal T}$ for which $A\cap T=\emptyset$ if and only if $B\cap T\neq\emptyset$.
Such $T$ is said to separate $A$ and $B$.
\end{definition}

Equivalently to Definition~\ref{def:trad_gt},  given a ground set $V=\{1,\ldots,n\}$,
a family of tests ${\cal T}=\{T_1,\ldots,T_k\}$ corresponds to a $k\times n$ matrix $T$ with $T_{ij}=1$ if $j \in T_i$, and $T_{ij}=0$ otherwise.
The outcome $y_{i,A}$ of the test $T_i$ on a subset $A$ is 1 if there exists $j\in A$ with $T_{ij}=1$, and 0 otherwise. Namely, $y_{i,A}=\vee_{j\in A}{T_{ij}}$.
With this notation, we seek a family of tests
${\cal T}=\{T_1,\ldots,T_k\}$ of minimum cardinality with corresponding matrix $T$ such that for any $A,B\subseteq V$ of size $d$ 
there exists $T_i\in {\cal T}$ for which $y_{i,A}\neq y_{i,B}$. 
Such $T_i$ is said to separate $A$ and $B$.
In traditional group testing, given a subset $S\subseteq V$ of contaminated items of cardinality $d$, the outcomes $y_{1,S},\ldots, y_{k,S}$ of \alex{tests} ${\cal T}=\{T_1,\ldots,T_k\}$ can be used to
reliably recover the contaminated subset $S$.

We now turn to define the object studied in this work - {\em generalized group testing} - in which one requires ${\cal T}=\{T_1,\ldots,T_k\}$ to separate not any two subsets $A$ and $B$ of size $d$, but rather any two subsets  $A$ and $B$ in a predetermined set system $E$.

\begin{definition}[generalized group testing (non-adaptive)]
\label{def:gen_gt}
Given a ground set $V$ of size $n$ and a family $E$ of subsets of $V$, find a minimum size family ${\cal T}$ of subsets of $V$ such that for any $A,B\in E$ there exists $T\in {\cal T}$ for which $A\cap T=\emptyset$ if and only if $B\cap T\neq\emptyset$.
\alexr{As before, such $T$ is said to separate $A$ and $B$. }
\end{definition}

Notice that the ground set $V$ and the family $E$ can be represented by a
 hypergraph $H=(V,E)$ whose vertices are the elements of the ground set and whose hyperedges are the sets in $E$.
As before, equivalently to Definition~\ref{def:gen_gt},
a family of tests ${\cal T}=\{T_1,\ldots,T_k\}$ corresponds to a $k\times n$ matrix $T$,  and in the generalized group testing problem we seek
a minimum sized family of tests ${\cal T}=\{T_1,\ldots,T_k\}$ with corresponding matrix $T$ such that for any $A,B\in E$ there exists $T_i\in {\cal T}$ for which $y_{i,A}\neq y_{i,B}$ (i.e., $T_i$ separates $A$ and $B$).
In the general group testing problem, for any \alex{possible}
subset $S \in E$ of contaminated items, the outcomes $y_{1,S},\ldots, y_{k,S}$ of ${\cal T}=\{T_1,\ldots,T_k\}$ can be used to
reliably recover $S$.

It is evident by our definitions that the traditional group testing problem with parameter $d$
corresponds to the generalized problem with a hypergraph $H=(V,E)$ in which the edge set consists of all subsets of $V$ of size $d$.

 We now turn to define the adaptive version of group testing in which one can design the tests ${\cal T}$ adaptively, that is, test $T_i$ may depend on the outcomes of tests $T_j$ for $j<i$.
 We present the definition for the generalized case, with the definition for the traditional setting following as a special case.

\begin{definition}[generalized adaptive group testing]
\label{def:gen_gt_ad}
Given a ground set $V$ of size $n$, a family $E$ of subsets of $V$, and a fixed but unknown subset $S\in E$ of contaminated items, 
interactively design a 
family ${\cal T}=\{T_1,\ldots,T_k\}$ of subsets of $V$ such that for any $2\le i\le k$
the choice of
$T_i$ 
depends on $\{y_{j,S} | j < i\}$, where 
$y_{j,S}$ 
is 1 if $S\cap T_j\neq \emptyset$, and 0 otherwise. The outcomes $y_{1,S},\ldots, y_{k,S}$ can be used to
reliably recover the contaminated subset $S$, in the sense that for any other $A \in E$ there exists an index $i$ such that $T_i$ separates $A$ and $S$, i.e., $y_{i,A} \ne y_{i,S}$.
The governing adaptive algorithm is said to use at most $k$ tests if for any $S\in E$, 
the interactively designed
 \alex{family}
of tests ${\cal T}$ is of size at most $k$.
One seeks to find an adaptive scheme \alex{that minimizes the value of $k$.}

\end{definition}

\subsection{Prior upper and lower bounds on group testing}
\label{sec:bounds}



We start by stating the information-theoretic lower bound that follows from the fact that each test (in both the adaptive and non-adaptive setting) yields at most one bit of information regarding the contaminated set (see, e.g., \cite{hl19,ajs19,dh00,j17}).

\begin{claim}\label{clm:ad_lb}(Adaptive $\&$ non-adaptive information-theoretic lower bound)
For any $0<d<n$, the size of an optimal \alex{adaptive or non-adaptive} solution for the traditional group testing problem with parameters $n$ and $d$ is  at least
$\Omega(d\log(n/d))$.
The size of an optimal solution for the generalized group testing problem with corresponding hypergraph $H=(V,E)$ is  at least
$\lceil\log_2|E|\rceil$.
\end{claim}

For non-adaptive traditional group-testing there is an improved lower bound \cite{efzf85}:
\begin{claim}\label{clm:non-ad_lb}(Non-adaptive lower bound)
For any $0<d<n$, the size of an optimal non-adaptive solution for the traditional group testing problem with parameters $n$ and $d$ is
$\Omega(\min\{n,d^2\log n/\log d\})$.
\end{claim}



Known upper-bounds for the traditional  group  testing  problem in the non-adaptive and adaptive setting (for general $n$, $d$) are given below (see e.g.,  \cite{dh00,pr08,ajs19}):



\begin{claim}\label{clm:non-ad_ub}(Non-adaptive upper bound)
For any $0<d<n$, all $d$ contaminated items in a given ground set $V$ of size $n$  
can be found using at most $O(d^2\log n)$ tests.

\end{claim}

\begin{claim} (Adaptive upper bound) 
For any $0<d<n$, all $d$ contaminated items in a given ground set of size $n$ can be found using at most $d \log_2 (n/d) + O(d)$ adaptive tests, even when $d$ is unknown.
\end{claim}

The table below compares the upper and lower bounds presented in this work (for small values of $d$) with those surveyed above (we use the notation: Adaptive (A), Non-Adaptive (NA), Upper-bound (UB), Lower-bound (LB)). Note that our results for generalized group testing on $H=(V,E)$ match, or come close to matching, those of traditional group testing when $E$ is taken to be all subsets of size $d$ in $V$ (and thus $\log{|E|}=\Theta(d\log(n/d))$).

\begin{center}
{\small
\begin{tabular}{c|c|c}
                & Traditional   & Generalized   \\
     \hline
     \alex{NA/UB}          &   $O(d^2\log n)$            &   $O(d \log{|E|})$            \\
\hline
    \alex{NA/LB}              &  $\Omega(d^2\log n/\log d)$             &    $\Omega(d\log {|E|}/\log d)$           \\
\hline
     \alex{A/UB}              &   $d \log_2 (n/d) + O(d)$            &  $O(\log{|E|}+d\log^2{d})$             \\
\hline
     \alex{A/LB}              &  $\Omega(d\log(n/d))$             &   $\Omega(\log{|E|} + d)$
     \end{tabular}}
\end{center}





\section{Non-adaptive Generalized Group Testing}
\subsection{The Computational Complexity of finding an Optimal or Approximately-Optimal Solution}

We first prove that the generalized group testing problem is NP-hard by showing a reduction from 3-colorability.
\begin{theorem}\label{thm:NP-hard}
Finding the size of an optimal solution for the (non-adaptive) generalized group testing problem is NP-hard.
\end{theorem}


\begin{prf}
Given a graph $G=(V,E)$, which is an instance of the 3-colorability problem, define the following instance of the generalized group testing problem. Let $V=\{1,2,\dots,n\}$ and assume without loss of generality  that $|E|=2^{\ell}-1$ for some integer $\ell \geq 2$ (otherwise one can modify $G$ by adding an additional 2-colorable component). Define the hypergraph $H=(V_H,E_H)$ for the generalized group testing problem as follows:
$V_H=E\cup V$ and
\begin{align*}
E_H =&
\left\{\{1,\dots, n,e\} \mid e\in E\right\} \\
& \cup
\left\{\{i,e\}| i\in V, e\in E, \exists j\in V\  \mbox{s.t}\  e=(i,j)\right\}.
\end{align*}
Note that $|V_H|=2^\ell-1+n$ (recall that $|E|=2^\ell-1$) and that $|E_H|=2^\ell-1+2(2^\ell-1)=2^{\ell+1}+2^\ell-3>2^{\ell+1}$. This latter fact implies (using the bounds of Claim~\ref{clm:ad_lb}) that the optimal solution for the generalized group testing problem corresponding to $H$ is of size at least $\ell+2$.

In what follows, we show that $G$ is 3-colorable (but not 2-colorable) if and only if the generalized group testing problem corresponding to $H$ can be solved using $\ell+2$ tests. 

For the first direction assume that $G$ is 3-colorable (but not 2-colorable). Define a $(\ell+2) \times (2^{\ell}-1+n)$ matrix $T$ as follows.
For every edge $e_m \in E$ let $u_m$ be a distinct identifying binary non-zero encoding using $\ell$ bits.
For $1\le m \le 2^{\ell}-1$ the $m$'th column in $T$ is an $(\ell+2)$-length vector whose first $\ell$ entries are $u_m$ and the last 2 entries are 0. For $2^{\ell} \le m \le 2^{\ell}+n-1$ the $m$'th column in $T$ is a $(\ell+2)$-length vector whose first $\ell$ entries are 0 and the last 2 entries are $00$ if for $i=m-(2^{\ell}-1)$, node $i \in V$ is colored with the first color, $01$ if $i$ is colored with  the second color, and $10$ if $i$ is colored with  the third color. The construction is illustrated in Figure~\ref{fig:reduction} on a simple example graph $G$ with 5 nodes.
\begin{figure}[t!]
    \begin{center}
    \vspace{-25mm}
           \includegraphics[scale=0.35]{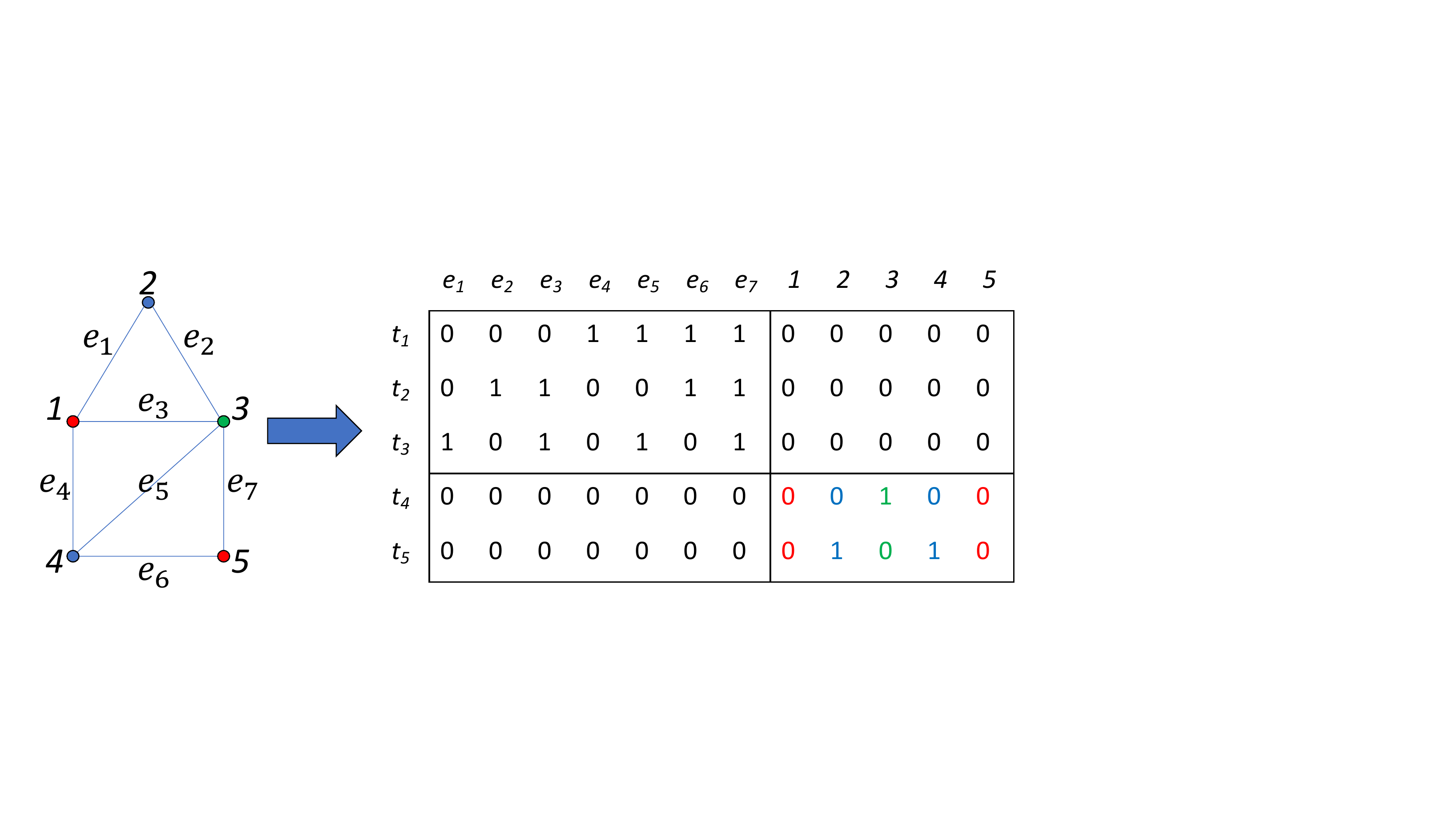}
    \vspace{-22mm}
    \caption{Illustration of the reduction in Theorem~\ref{thm:NP-hard}. As $|V|=7$, the parameter $\ell$ equals $3$. Note that each edge $e_m$ has a unique vector $u_m$ encoded in rows $t_1, t_2, t_3$, and each vertex has a unique color encoded by rows $t_3, t_4$.}\label{fig:reduction}
    \end{center}
\end{figure}

To show that $T$ is a feasible solution for the generalized group testing problem we prove that for any hyperedges $A,A'\in E_H$ there exists $1\le t \le \ell+2$ such that $y_{t,A}\neq y_{t,A'}$.
First assume that $A$ and $A'$ satisfy one of the following cases for $m \ne m'$: $A=\{1,\ldots,n,e_m\}$, $A'=\{1,\ldots,n,e_{m'}\}$;
or $A=\{i,e_m\}$, $A'=\{i',e_{m'}\}$;
or $A=\{i,e_m\}$, $A'=\{1,\ldots,n,e_{m'}\}$. In these cases, since the encodings $u_m$ of $e_m$ and $u_{m'}$ of $e_{m'}$ are distinct, there exists an entry $t$, $1\le t\le \ell$, for which $T_{t,m}\neq T_{t,m'}$. This, in turn, implies (given the construction of $T$) that $y_{t,A}=\vee_{j\in A}{T_{t,j}} \ne \vee_{j\in A'}{T_{t,j}}=y_{t,A'}$.

If $A=\{1,\ldots,n,e_m\}$ and $A'=\{i',e_{m}\}$, then, by our construction of $T$ there exists $t\in \{\ell+1,\ell+2\}$ for which $T_{t,2^{\ell}-1+i'}=0$. For such $t$, there must be an $r \in V$ for which $T_{t,2^{\ell}-1+r}=1$, as otherwise the nodes in $V$ are colored by two colors alone. Thus, it is not hard to verify that in this case as well $y_{t,A}\neq y_{t,A'}$.

Finally, if $A=\{i,e_m\}$, $A'=\{i',e_m\}$ where $e_m=(i,i')$, then since $i$ and $i'$ are assigned to distinct colors in $G$ we get that there exists $t\in \{\ell+1,\ell+2\}$ such that $T_{t,2^{\ell}-1+i}\neq T_{t,2^{\ell}-1+i'}$. As above, in this case as well $y_{t,A}\neq y_{t,A'}$. Therefore any two subsets of $E_H$ can be separated.

For the second direction, assume that ${\cal T}$ is a solution for the generalized group testing problem with $|{\cal T}|=\ell+2$, and let $T$ be the corresponding $(\ell+2)\times (2^{\ell}-1+n)$ matrix. We show that $G$ is 3-colorable. Denote by $T_1^{col},\ldots,T_{2^{\ell}-1+n}^{col}$  the columns of $T$, and let $w$ be the vector corresponding to the union of the last $n$ columns in $T$.
Namely $w=T_{2^{\ell}}^{col}\vee,\dots,\vee T_{2^{\ell}-1+n}^{col}$ and $w_t=1$ if and only if there exists $m \ge 2^\ell$ for which $T_{t,m}=1$.
Then, it holds that the support $|{\tt sup}(w)|$ of $w$ is at most of size 2. Assume to the contrary that $|{\tt sup}(w)|\ge 3$, and without loss of generality that  $w_{\ell}=w_{\ell+1}=w_{\ell+2}=1$.
As there are $2^\ell-1$ edges $e_m$ in $E$, we have that tests $T_1$ up to $T_{\ell-1}$ cannot separate at least one pair of subsets $A,A'$ in $E_H$ of the form $\{1,\ldots, n,e_m\}$ and $\{1,\ldots, n,e_{m'}\}$. This fact follows from the lower-bound of $\log_2|E|$ given in Claim~\ref{clm:ad_lb}. Moreover, these same subsets $A$ and $A'$ cannot be separated by tests $T_{\ell},T_{\ell+1}$, and $T_{\ell+2}$ by our assumption that $w_{\ell}=w_{\ell+1}=w_{\ell+2}=1$. Thus $T$ is not a feasible solution to the generalized group testing problem corresponding to $H$, a contradiction.

We can now assume that $|{\tt sup}(w)|\le 2$. Without loss of generality assume that entries 1 to $\ell$ of $w$ are 0. Then for every $1\le i\le n$ it holds that $|{\tt sup}(T_{2^{\ell}-1+i}^{col})|<2$, as otherwise both $T_{\ell+1,2^{\ell}-1+i}=1$ and $T_{\ell+2,2^{\ell}-1+i}=1$ implying that for any $e=(i,j) \in E$ the subsets $A=\{1\ldots,n,e\}$ and $A'=\{i,e\}$ can not be separated.

For a coloring of $G$, assign to node $i$ in $V$ the color corresponding to the values in $T_{\ell+1,2^{\ell}-1+i},T_{\ell+2,2^{\ell}-1+i}$. This implies a legal coloring of $G$ using the colors 00,01,10. Specifically, if $e=(i,j)$ then there exists a test $T_t\in {\cal T}$ that separates $A=\{i,e\}$ and $A'=\{j,e\}$ implying that  $T_{t,2^{\ell}-1+i}\neq T_{t,2^{\ell}-1+j}$, which in turn guarantees that $i$ and $j$ are assigned to distinct colors.
\end{prf}

We next extend Theorem~\ref{thm:NP-hard} to \alex{establish the hardness of  approximation of the size of an optimal group testing solution.}
Namely,
we address the question of approximating the size of the optimal solution for the generalized group testing problem within a multiplicative factor of $1+\varepsilon$ (for small $\varepsilon$). We again reduce from 3-colorability, using the fact that  coloring a $3$-colorable graph with $n^\varepsilon$ colors (for sufficiently small $\varepsilon$) is a well known open problem (see e.g. \cite{KLS00,langberg2008graph} and references therein).
Our proof 
closely follows that of Theorem~\ref{thm:NP-hard}.




\begin{theorem}\label{thm:hardness-apprx}
Let $\varepsilon>0$ be sufficiently small.
Approximating the size of an optimal solution for the (non-adaptive) generalized group testing problem within a multiplicative factor of $1+\varepsilon$ is as hard as coloring a $3$-colorable graph with at most $n^{4\varepsilon}$ colors.
\end{theorem}


\begin{prf}
Given a graph $G=(V,E)$, which is an instance of the 3-colorability problem, define as in the proof of Theorem~\ref{thm:NP-hard} the corresponding instance $H=(V_H,E_H)$ of the generalized group testing problem. 
By the proof of Theorem~\ref{thm:NP-hard}, if $G$ is 3-colorable (and not 2-colorable) then the generalized group testing problem corresponding to $H$  can be solved (optimally) using $\ell+2$ tests.
 We now show that if one can find a solution to the generalized group testing problem corresponding to $H$ using at most  $(\ell+2)(1+\varepsilon)$ tests, then one can color $G$ with at most $n^{4\varepsilon}$ colors.




Let $\Delta=\ell\varepsilon+2(1+\varepsilon)$.
Then, since $\ell+\Delta=(\ell+2)(1+\varepsilon)$, there is a solution ${\cal T}$ for the generalized group testing problem with $|{\cal T}|=\ell+\Delta$.
Let $T$ be the corresponding $(\ell+\Delta)\times (2^{\ell}-1+n)$ matrix. We show that $G$ is $(2^{\Delta}-1)$-colorable. As in the proof of Theorem~\ref{thm:NP-hard}, denote the columns of $T$ by $T_1^{col},\ldots,T_{2^{\ell}-1+n}^{col}$, and let $w=T_{2^{\ell}}^{col}\vee,\ldots,\vee T_{2^{\ell}-1+n}^{col}$.
Then it holds that $|{\tt sup}(w)|\le \Delta$. Assume to the contrary that $|{\tt sup}(w)|\ge \Delta+1$. Without loss of generality assume that the $\ell,\ell+1,\ldots,\ell+\Delta$'th entries of  $w$ are 1.
As there are $2^\ell-1$ edges $e_m$ in $E$, we have that tests $T_1$ up to $T_{\ell-1}$ cannot separate at least one pair of subsets $A,A'$ in $E_H$ of the form $\{1,\ldots, n,e_m\}$ and $\{1,\ldots, n,e_{m'}\}$. This fact follows from the lower-bound of $\log_2|E|>\ell-1$ given in Claim~\ref{clm:ad_lb}. Moreover, these same subsets $A$ and $A'$ cannot be separated by tests $T_{\ell},T_{\ell+1}, \dots, T_{\ell+\Delta}$ by our assumption that $w_{\ell}=w_{\ell+1}=\dots=w_{\ell+\Delta}=1$. Thus $T$ is not a feasible solution to the generalized group testing problem corresponding to $H$, a contradiction.


Therefore $|{\tt sup}(w)|\le \Delta$.  Without loss of generality assume that entries 1 to $\ell$ of $w$ are 0.
We now assign to vertex $i$ in $G$ the color corresponding to the binary vector $T_{\ell+1,2^{\ell}-1+i},\ldots,T_{\ell+\Delta,2^{\ell}-1+i}$. The coloring is  valid.
Namely, for any $e=(i,j)$ there exists a test $T_t\in {\cal T}$ that separates $A=\{i,e\}$ and $A'=\{j,e\}$ implying that  $T_{t,2^{\ell}-1+i}\neq T_{t,2^{\ell}-1+j}$, which in turn guarantees that $i$ and $j$ are assigned to distinct colors.
To conclude, we note that the total number of colors used in the coloring of $G$ is at most $2^{\Delta}=2^{\ell\varepsilon+2(1+\varepsilon)}\le 2^{(2\log_2 n)\varepsilon+2(1+\varepsilon)}<n^{4\varepsilon}$ (for sufficiently large $n$). Here, we use the fact that as $2^{\ell}-1=|E|\leq {n \choose 2}$, it follows that $\ell \leq 2\log_2n$.
\end{prf}

\subsection{Non-adaptive generalized group testing bounds}


In the following theorem we present an upper bound for the non-adaptive generalized group testing problem. The stated bound is a function of the set-system size $|E|$, the maximum size $d$ of any $e \in E$, and a parameter $\beta$ which addresses the maximum pair-wise symmetric difference size of any two subsets in $E$. Loosely speaking, symmetric difference is an important and natural primitive, since if  $|e \setminus e'|$ or $|e' \setminus e|$ is {\em large}, then it is {\em easier} for a test to separate $e$ from $e'$. Implying that less tests may be needed when such sizes are large. Using a rough analog from coding theory, one can view the tests ${\cal T}$ as a syndrome based decoding process, and the (indicator vector of) edges $e$ as codewords. In this analog, $\max\{|e \setminus e'|, |e' \setminus e|\}$ is related to distance, implying intuitively that a collection of codewords with large minimum distance is easier to decode than a collection with small minimum distance.
In what follows, we set
$\beta=\min_{e,e' \in E}\max\{|e \setminus e'|, |e' \setminus e|\}$.
For  $\beta=\Theta(d)$ our solution size matches the information-theoretic lower bound of $\log_2 |E|$ stated in Claim~\ref{clm:ad_lb} (up to a constant multiplicative factor).
In the case in which no assumptions are made on $\beta$, our solution of size $O(d\log_2 |E|)$ comes close to matching our (worst-case) lower bound of $\Omega(d\log{|E|}/\log{d})$.
Our solution is generated by constructing tests at random, refining the analysis appearing in, e.g., \cite{ams06},  addressing traditional group testing.

\begin{theorem}
\label{thm:upper_na}
Consider an instance of the generalized group testing problem with corresponding hypergraph $H=(V,E)$ for which all edges are of size at most $d$.
In addition, assume that for all $e,e'\in E$ it holds that $\max\{|e \setminus e'|, |e' \setminus e|\} \ge \beta$, for some parameter $1\le\beta<d$. 
Then there exists a solution ${\cal T}$ to the generalized group testing problem corresponding to $H$ of size $O(\frac{d}{\beta}\log |E|)$.
Moreover, for any constant $\alpha$, there exists an efficient randomized construction of ${\cal T}$ of size $O(\frac{d}{\beta}(\log |E|+\alpha))$ that is a valid solution with probability at least $1-e^{-\alpha}$.
\end{theorem}

Theorem~\ref{thm:upper_na} immediately implies the following corollary in which no restrictions are posed on the intersection size of edges in $H$ (namely, the parameter $\beta$ of Theorem~\ref{thm:upper_na} equals 1).

\begin{coro}
\label{cor:upper_na}
Consider an instance of the generalized group testing problem with corresponding hypergraph $H=(V,E)$ for which all edges are of size at most $d$.
Then there exists a solution ${\cal T}$ of size $O(d\log |E|)$.
\end{coro}

\begin{prf} {\em (of Theorem~\ref{thm:upper_na})}
Our proof follows that appearing in, e.g., \cite{ams06} addressing the traditional group testing problem.
Let $|V|=n$.
Consider a collection of $k$ tests ${\cal T}=\{T_1,\dots,T_k\}$ represented by the $k \times n$ matrix $T$ in which each entry of $T$ is independently set to be 1 with probability $p=1/d$ (and 0 otherwise).
Consider two edges $e, e'\in E$.
Assume, without loss of generality, that $|e \setminus e'| \ge \beta$.
Let $T_i$ be the test (i.e., subset) corresponding to the $i$'th row of $T$.
The probability (over $T_i$) that $e$ and $e'$ are separated by $T_i$ is at least
\begin{align*}
\Pr[T_i \cap & e\ne \emptyset,\
T_i \cap e' = \emptyset] \\
& = \Pr[T_i \cap  (e\setminus e') \ne \emptyset,\
T_i \cap e' = \emptyset] \\
& \ge (1-p)^d\left(1-(1-p)^{\beta}\right) \\
& = \left(1-\frac{1}{d}\right)^d\left(1-\left(1-\frac{1}{d}\right)^{\beta}\right) \\
& \ge \frac{1}{e}\left(1-\left(1-\frac{1}{d}\right)^{\beta}\right)
\end{align*}

Therefore, using the union bound, we get that
\begin{align*}
\Pr_T[\exists & e,e'\in E\ \mbox{that are not separated by $k$ tests}]\\
& \le {|E|\choose 2}\cdot \left(1-\frac{1}{e}\left(1-\left(1-\frac{1}{d}\right)^{\beta}\right)\right)^k
\end{align*}
Let $\alpha$ be a large constant.
Thus, for
$$
k>-\frac{2\ln|E|+\alpha}{\ln\left(1-\frac{1}{e}\left(1-\left(1-\frac{1}{d}\right)^{\beta}\right)\right)}
$$
we get that
$$
\Pr_T[\exists e,e'\in E\ \mbox{that are not separated by ${\cal T}$}]<e^{-\alpha}.
$$






Using the fact that for $x\in[0,1/2]$, $x \le -\ln(1-x) \le 2x$ we have
\begin{align*}
    -\ln  \left(1-\frac{1}{e}\left(1-\left(1-\frac{1}{d}\right)^{\beta}\right)\right)
    & \ge \frac{1}{e}\left(1-\left(1-\frac{1}{d}\right)^{\beta}\right),
\end{align*}
and $\beta\ln\left(1-\frac{1}{d}\right) \leq -\beta\frac{1}{d} \leq \ln\left(1-\frac{\beta}{2d}\right)$ which implies
$\left(1-\frac{1}{d}\right)^{\beta} \leq 1-\frac{\beta}{2d}$.
Thus,
\begin{align*}
    -\ln & \left(1-\frac{1}{e}\left(1-\left(1-\frac{1}{d}\right)^{\beta}\right)\right)  \ge \\
    & \ge \frac{1}{e}\left(1-\left(1-\frac{1}{d}\right)^{\beta}\right) \ge \frac{1}{e}\left(1-\left(1-\frac{\beta}{2d}\right)\right) \\
    & = \frac{\beta}{2ed}
\end{align*}

Therefore $\frac{2\ln |E| +  \alpha}{-\ln\left(1-\frac{1}{e}\left(1-(1-\frac{1}{d})^{\beta}\right)\right)}\le (2\ln |E|+ \alpha)\frac{2ed}{\beta}$, so for $k =(2\ln |E|+ \alpha)\frac{2ed}{\beta}$ it holds that $
k>-\frac{2\ln|E|+\alpha}{\ln\left(1-\frac{1}{e}\left(1-\left(1-\frac{1}{d}\right)^{\beta}\right)\right)}
$ and in turn $\Pr[\exists e,e'\in E$ that are not separated by ${\cal T}]<e^{-\alpha}$.

Setting $\alpha=1$ concludes our proof for the existence of ${\cal T}$ of size $O(\frac{d}{\beta}\ln |E|)$, while taking $\alpha$ to be large allows an efficient randomized construction of ${\cal T}$ that is a valid solution with probability at least $1-e^{-\alpha}$.
\end{prf}

To complete our analysis of the non-adaptive setting we present a lower-bound to Corollary~\ref{cor:upper_na}.
Our bound strongly builds on, and generalizes, Claim~\ref{clm:non-ad_lb} which corresponds to traditional group testing.

\begin{claim}
\label{claim:non_adap}
Let $d \leq n' \leq n$ 
be integers.
There exist hypergraphs $H=(V,E)$ with $|V|=n$, edges of size $d$, and $|E|={n' \choose d}$ such that any non-adaptive solution for the generalized group testing problem corresponding to $H$ requires $\Omega(\min\{n',d\log{|E|}/\log{d}\})= \Omega(\min\{n',d^2\log{n'}/\log{d}\})$ tests.
\end{claim}

\begin{prf}
Consider the hypergraph $H$ with vertex set $V=V' \cup U$ where $|V'|=n'$ and $|U|=n-n'$, and edge set $E$ which includes all $d$ sized subsets of $|V'|$ (and nothing else).
On one hand, it holds by Claim~\ref{clm:non-ad_lb} that any non-adaptive solution for the traditional group testing problem corresponding to $H$ requires at least $\Omega(\min\{n',d^2\log{n'}/\log{d}\})$ tests.
On the other hand, we have that
$\log{|E|}=\log{n'\choose d} \leq  2d\log(n'/d)$.
Thus, any non-adaptive solution for the generalized group testing problem corresponding to $H$ requires
$\Omega(\min\{n',d\log{|E|}/\log{d}\})= \Omega(\min\{n',d^2\log{n'}/\log{d}\})$ tests for $|E|={{n'} \choose d}$.
\end{prf}
\section{Adaptive generalized group testing}

We now address the adaptive setting.
Roughly speaking, adaptive schemes improve on non-adaptive constructions as the adaptive iterative process allows to {\em rule-out} potential contaminated items or subsets of the population in each iteration.
While, in traditional group testing, the adaptive analysis is commonly governed by the {\em ground-set} size of potentially contaminated individuals, which shrinks with each iteration of the iterative process (see, e.g. \cite{dh00,bja13}),
the analysis in our generalized setting must be governed by the number of remaining {\em edges} that are potentially contaminated. This difference causes a number of challenges that are addressed in the proof below.

\begin{theorem}\label{thm:adaptive}
Consider an instance $H=(V,E)$ of the generalized group testing problem in which edges in $E$ are of size  $d$. 
There is an adaptive algorithm that
interactively designs a collection of tests ${\cal T}$ of size at most $O(\log |E|+d\log^2 d)$ that can reliably recover the contaminated set in $E$.
Moreover, there exists instances $H=(V,E)$ with corresponding bound $d$ for which any adaptive solution is of size at least $\Omega(\log{|E|}+d)$.
\end{theorem}

\begin{prf}
We first address the lower bound of $\Omega(\log{|E|}+d)$. Consider the instance $H=(V,E)$ where $V=\{1,2,\dots,d+1\}$ (that is, $|V|=d+1$) and $E$ consists of all subsets of $V$ of size $d$. Namely, every edge $e$ in $E$ includes all but one node in $V$.

Consider a first test $T \subset V$ in a given adaptive scheme. Observe that if $T$ includes more than a single node it will always be positive (as any subset of size $2$ will intersect all edges in $E$). Thus, the only test $T$ that will add information is a test that includes a single node. Without loss of generality, assume that $T=\{1\}$.
A negative outcome after testing with $T$ implies that the contaminated edge is $e=V \setminus \{1\}$. Otherwise, we learn only that edge $e=V \setminus \{1\}$ in not contaminated as all remaining edges include node $\{1\}$. Thus, the instance is reduced to one in which $V'=\{2,3,\dots,d+1\}$ (that is, $|V'|=d$) with $|E'|$ that  includes all edges of size $d-1$ in $V'$.
Continuing with the new instance, we again notice that any test must include at most a single node from $V'$ and thus if negative finds the contaminated edge, and if positive reduced the instance size by 1. We apply the same argument for each additional test.

As any adaptive algorithm is characterized by the sequence of tests it performs, which in turn, for the instance at hand, are characterized by an ordering of the vertices $V=\{1,2,\dots,d+1\}$.
We conclude that for any such deterministic (randomized) ordering, the worst-case (expected) number of tests needed is bounded by below by $d$ ($\Omega(d)$). As $|E|=d+1$, we conclude that the number of tests required in any adaptive algorithm for the instance at hand is at least $\Omega(\log{|E|}+d)$.

For the upper bound, let $V=\{1,2,\dots,n\}$ (that is, $|V|=n$), and for $1\le i\le n$, let $d_i$ be the degree of node $i$ in $V$. Notice that $\sum_{i=1}^n d_i=d\cdot |E|$.
In what follows, we present an adaptive algorithm using $O(\log |E|+d\log^2 d)$ tests.
Roughly speaking, the algorithm proceeds in rounds in which we search for subsets $T \subseteq V$ that intersect a constant fraction of the edge set. In round $j$, the constant fraction is required to be in the range $[\varepsilon_j,1-\varepsilon_j]$, for $\varepsilon_j=1/2^{j+1}$. Once such a subset $T$ is found and used as a test, we are able to reduce the edge size of the instance at hand by a factor if
$(1-\varepsilon_j)$ and thus to {\em make progress} towards finding the contaminated $e \in E$.

Each round $j$ consists of several sub-rounds in which one consecutively finds subsets $T$ corresponding to the same parameter $\varepsilon_j$, thus reducing the size of the edge set in each sub-round by a factor of $(1-\varepsilon_j)$. We move from round $j$ to round $j+1$ once no such $T$ corresponding to $\varepsilon_j$ is found (and thus increase $j$ by 1 allowing for additional flexibility in the requirements on $T$).
Finally, once $\varepsilon$ reaches $1/d$, we stop the iterative process and solve the remaining instance in a non-adaptive manner.
The algorithm is presented below:

\vspace{-2mm}
\BA\label{alg:adaptive}{{\bf (Adaptive algorithm)}}
\begin{enumerate}
\item $E'\longleftarrow E$, $\varepsilon\longleftarrow 1/4$.
\item \label{item:loop} If $|E'|=1$, return $E'$.
\item Label the nodes in $V$ such that $d_1\le d_2\le\ldots\le d_n$ where $d_i$ be the degree of node $i$ in $V$ w.r.t. $E'$.
  \item \label{item:find_t}If there exists a subset $T\subseteq V$ of the form $T=\{1,\ldots t\}$ (i.e., $T$ takes nodes in increasing order of degree) such that the number of edges of $E'$ intersecting $T$ is in the range $[\varepsilon|E'|,(1-\varepsilon)|E'|]$ then:
    \begin{enumerate}
      \item \label{item:test} \alex{Perform test with subset $T$.}
      \begin{enumerate}
          \item  If the test is negative, set $E'\longleftarrow \{e\in E'|e\cap T=\emptyset\}$ (i.e., update $E'$ to  the subset of edges that do not intersect $T$),
          and return to (\ref{item:loop}).
          \item If the test is positive, then set $E'\longleftarrow \{e\in E'|e\cap T\neq\emptyset\}$ (i.e., update $E'$ to be the subset of edges that do intersect $T$) and return to (\ref{item:loop}).
      \end{enumerate}
  \item \label{item:final} Else, if no such $T$ is found, if $\varepsilon>1/d$ then return to (\ref{item:loop}) with $\varepsilon\longleftarrow \varepsilon/2$. Otherwise, continue with $E'$, using the non-adaptive testing described in Theorem~\ref{thm:upper_na} to find the contaminated $e \in E$.
  \end{enumerate}
\end{enumerate}
\EA

We first address the correctness of the proposed algorithm.
Let $e \in E$ be the contaminated subset. It suffices to show that throughout the execution of the adaptive algorithm, the subset $e$ is in $E'$. This holds initially as $E'=E$, and continues to hold in Step~\ref{item:test} by the fact that $T$ is positive if and only if $e \cap T \ne \emptyset$.

We now compute the number of tests performed by the algorithm.
Consider round $j$ in which $\varepsilon=\varepsilon_j=1/2^{j+1}$.
We first show that if $T$ is not found in Step~\ref{item:find_t} then it must be the case that $E'$ is of size at most $(1+2\varepsilon_j) 2^{d(1+4\varepsilon_j) H(2\varepsilon_j)}$.
Specifically, if there is no test $T$ as described in Step~\ref{item:find_t}, then it must be the case that for some node $i$ the number of edges intersecting $\{1,2,\ldots i-1\}$ is less than $\varepsilon_j |E'|$ and  the number of edges intersecting $\{1,2,\ldots i\}$ is more than $(1-\varepsilon_j) |E'|$. This implies that $d_i\ge (1-\varepsilon_j) |E'| -\varepsilon_j |E'| = (1-2\varepsilon_j) |E'|$. As $d_i\le d_{i+1} \le \ldots \le d_n$, and $\sum_{\ell=1}^n{d_{\ell}}=d|E'|$ it holds that $(n-i+1)(1-2\varepsilon_j)|E'| \le \sum_{\ell=i}^n{d_{\ell}}\le d|E'|$. Therefore, $(n-i+1)(1-2\varepsilon_j)\le d$, so $n-i+1 \le d/(1-2\varepsilon_j)$.

Recall that less than $\varepsilon_j |E'|$ edges are adjacent to nodes $\{1,\ldots, i-1\}$.
This implies that at least $(1-\varepsilon_j)|E'|$ edges are induced by nodes $\{i,\ldots,n\}$. As $|\{i,\ldots,n\}| = n-i+1\le  d/(1-2\varepsilon_j)$, the number of edges of size $d$ induced by $\{i,\ldots,n\}$ is at most ${d/(1-2\varepsilon_j) \choose d}$.
This implies that $(1-\varepsilon_j)|E'|\le {d/(1-2\varepsilon_j) \choose d} =
{d/(1-2\varepsilon_j) \choose
2\varepsilon_j \cdot d/(1-2\varepsilon_j)}$, so
$
|E'|\le \frac{2^{d\cdot H(2\varepsilon_j)/(1-2\varepsilon_j)}}{1-\varepsilon_j}\leq
(1+2\varepsilon_j)2^{d(1+4\varepsilon_j) H(2\varepsilon_j)}.
$

To bound the number of tests in the adaptive algorithm, it suffices to analyze the total number of sub-rounds executed.
Let $m_j$ be the size of $E'$ in the beginning of round $j$.
From the above, we have that $m_1=|E|$, and $m_j \leq (1+2\varepsilon_j) 2^{d(1+4\varepsilon_j) H(2\varepsilon_j)}$, for any $j=1,2,\dots,\log{d}$.
Moreover, as the size of $E'$ in any sub-round of round $j$ reduces its size by a factor of $1-\varepsilon_j \leq \frac{1}{1+\varepsilon_j}$.
The total number of sub-rounds is thus bounded by
{\small
\begin{align*}
\sum_{j=1}^{\log{d}}&\frac{\log(m_j)-\log(m_{j+1})}{\log{(1+\varepsilon_j)}}  \leq
\sum_{j=1}^{\log{d}}(\log(m_j)-\log(m_{j+1}))\frac{1}{\varepsilon_j} \\
& \leq
\frac{\log{m_1}}{\varepsilon_1}+\sum_{j=2}^{\log{d}}\log(m_j)\cdot \left(\frac{1}{\varepsilon_{j}}-\frac{1}{\varepsilon_{j-1}}\right) \\
& =
4\log{|E|}+\sum_{j=2}^{\log{d}}\log(m_j)\cdot 2^{j-1}\\
& \leq
4\log{|E|}+\sum_{j=2}^{\log{d}}\log\left((1+2\varepsilon_j) 2^{d(1+4\varepsilon_j)H(2\varepsilon_j)}\right)\cdot 2^{j-1}\\
& =
O\left(\log{|E|}+\sum_{j=2}^{\log{d}}d H(2\varepsilon_j)\cdot 2^{j}\right)\\
& =
O\left(\log{|E|}+\sum_{j=2}^{\log{d}}d \cdot 2\varepsilon_j\log(1/2\varepsilon_j)\cdot 2^{j}\right)\\
& =
O\left(\log{|E|}+\sum_{j=2}^{\log{d}}d \cdot j\right)
 =
O\left(\log{|E|}+d\log^2{d}\right).
\end{align*}}


Finally, to bound the total number of tests in the suggested algorithm, we add the number of tests performed in step~\ref{item:final} for the set $E'$ once $\varepsilon_j \le 1/d$.
At this stage, the set $E'$ is of size at most $(1+2\varepsilon_j) 2^{d(1+4\varepsilon_j) H(2\varepsilon_j)} \le 2^{O(d \cdot \frac{1}{d}\log{d})}=d^{O(1)}$.
Thus, the non-adaptive testing described in Theorem~\ref{thm:upper_na} 
when applied to $E'$ will use at most
$O(d\log{|E'|}) = O(d\log{d})$ tests.
We conclude that, all in all, the suggested adaptive algorithm uses
$O\left(\log{|E|}+d\log^2{d}\right)$ tests.
\end{prf}

We remark that an adaptive upper-bound of $O(\log{|E|}+d^2)$ holds, using a similar proof, for the modified setting in which edges of $E$ are of size at most $d$ (as apposed to exactly $d$).

\section{Concluding remarks}
\label{sec:diss}

In this paper we consider a generalization of the traditional group testing problem
in which the contaminated set is one of a  collection of subsets characterized by the edge set of a hypergraph $H=(V,E)$. 
Leveraging this additional knowledge, we address both adaptive and non-adaptive group testing, in terms of upper and lower bounds, and for the latter, analyze the complexity of determining the size of optimal (or approximately optimal) solutions.

Beyond adaptive and non-adaptive group testing, several additional models have been studied for traditional group testing.
These include, noisy group testing, probabilistic group testing, partial recovery, non-binary outcomes, 
an unknown bound on $d$, and more. Addressing those models in our generalized setting is  subject to future research.


\newpage

\end{document}